\documentclass[11pt]{article}

\usepackage[preprint]{acl}

\usepackage{times}
\usepackage{latexsym}
\usepackage{hyperref}
\usepackage{url}
\usepackage{graphicx}
\usepackage{subcaption}
\usepackage{booktabs}
\usepackage[T1]{fontenc}

\usepackage[utf8]{inputenc}

\usepackage{microtype}

\usepackage{inconsolata}

\usepackage{graphicx}

\title{Quantized Inference for OneRec-V2}

\author{Yi Su \\\And
  Xinchen Luo \\\And
  Hongtao Chen \\\And
  Ziteng Shu \\\And
  Liang Zeng \\\And
  Yunfeng Zhao \\\And
  Fangyu Zhang \\\And
  Jiaqiang Liu \\\And
  Xiao Liang \\\And
  Yiwu Liu \\\And
  Ruiming Tang}

\author{
 \textbf{Yi Su},
 \textbf{Xinchen Luo},
 \textbf{Hongtao Cheng},
 \textbf{Ziteng Shu},
 \textbf{Yunfeng Zhao},
\\
 \textbf{Fangyu Zhang},
 \textbf{Jiaqiang Liu},
 \textbf{Xiao Liang},
 \textbf{Yiwu Liu},
 \textbf{Ruiming Tang}
\\
 Kuaishou Inc., Beijing, China
\\
 tangruiming@kuaishou.com
}

\begin{document}
\maketitle
\begin{abstract}
Quantized inference has demonstrated substantial system-level benefits in large language models while preserving model quality. In contrast, reliably applying low-precision quantization to recommender systems remains challenging in industrial settings. This difficulty arises from differences in training paradigms, architectural patterns, and computational characteristics, which lead to distinct numerical behaviors in weights and activations. Traditional recommender models often exhibit high-magnitude and high-variance weights and activations, making them more sensitive to quantization-induced perturbations. In addition, recommendation workloads frequently suffer from limited hardware utilization, limiting the practical gains of low-precision computation.
In this work, we revisit low-precision inference in the context of generative recommendation. Through empirical distribution analysis, we show that the weight and activation statistics of OneRec-V2 are significantly more controlled and closer to those of large language models than traditional recommendation models. Moreover, OneRec-V2 exhibits a more compute-intensive inference pattern with substantially higher hardware utilization, enabling more end-to-end throughput gains with low-precision computation. Leveraging this property, we develop a FP8 post training quantization framework and integrate it into an optimized inference infrastructure. The proposed joint optimization achieves a 49\% reduction in end-to-end inference latency and a 92\% increase in throughput. Extensive online A/B testing further confirms that FP8 inference introduces no degradation in core metrics.
These results suggest that as recommender systems evolve toward the paradigms of large language models, algorithm-level and system-level optimization techniques established in the LLM domain can be effectively adapted to large-scale recommendation workloads.
\end{abstract}
\section{Introduction}

Quantized inference has become an essential technique for improving the efficiency of large-scale neural networks. In large language models (LLMs), low-precision formats have demonstrated substantial system-level benefits while preserving model quality \citep{fp8,smoothquant,llm-int8,mixed-precision-training}. These successes are closely tied to the dense, compute-intensive execution patterns and relatively well-regulated numerical behaviors of Transformer-based architectures.

In contrast, applying low-precision quantization reliably in recommender systems has long remained challenging in industrial practice. Traditional recommendation models are typically optimized for fine-grained ranking tasks and differ significantly from LLMs in both training paradigms and architectural structures \citep{reco-system,scale-reco,review-reco}. Empirically, their weights and activations often exhibit high magnitudes and large variances, which make these models more sensitive to quantization-induced perturbations. From a systems perspective, classical recommender inference workloads are frequently memory or control bound and exhibit relatively low hardware utilization \citep{reco-system,torchrec,onerec}. As a result, even when hardware platforms support low-precision computation, the practical end-to-end gains may be limited. These numerical and system factors have historically hindered the effective deployment of low-precision inference in traditional recommendation pipelines.

Recent advances in generative recommendation models have begun to narrow this gap. OneRec introduces a unified generative framework that integrates retrieval and ranking \citep{onerec}, and subsequent extensions such as OneRec-V2 further refine this paradigm through architectural scaling and training improvements \citep{onerec-v2,onerec-think,openonerec}. Compared to classical recommendation architectures, these models rely more heavily on dense and structurally constrained computation paths, increasing computation intensity and improving hardware utilization during inference. At the same time, their training paradigms impose stronger implicit regularization on activation magnitudes and variances.

In this work, we investigate quantized inference for OneRec-V2. Through empirical distribution analysis, we show that the weight and activation statistics of OneRec-V2 are significantly more controlled and closer to those of LLMs than those of traditional recommendation models. Moreover, its compute-intensive inference pattern allows low-precision computation to translate more effectively into end-to-end system improvements. Leveraging these properties, we develop a FP8 post training quantization framework integrated with an optimized inference infrastructure.
Extensive experiments demonstrate that the proposed system achieves a 49\% reduction in end-to-end inference latency and a 92\% increase in throughput. Online A/B testing further confirms that our quantized inference introduces no degradation in core metrics. These results suggest that as recommender systems evolve toward the paradigms of large language models, low-precision techniques established in the LLM domain can be safely and effectively adapted to modern recommendation workloads.

The main contributions of this paper are summarized as follows:

\begin{itemize}
  \item We analyze the numerical and system factors that have historically limited the effectiveness of low-precision inference in traditional recommender systems.
  \item We empirically demonstrate that the architectural and training paradigm shifts in OneRec-V2 lead to more controlled weight and activation statistics and higher hardware utilization, substantially improving the feasibility of quantized inference.
  \item We develop a FP8 post training quantization framework integrated with an optimized inference infrastructure, achieving significant latency reduction and throughput improvement without any degradation in online performance.
\end{itemize}
\section{Related Work}
\subsection{Low-Precision Quantization}
Low-precision quantization aims to reduce computation and memory costs by representing weights and activations with lower numerical precision while controlling the resulting approximation error. In recent years, extensive studies have demonstrated that low-precision quantization can be applied effectively to large language models, achieving substantial inference acceleration with minimal quality degradation \citep{llm-int8,smoothquant,quantization1,quantization2,quantization3}. 

These advances highlight the importance of aligning quantization strategies with model architecture and activation statistics. When numerical distributions are well regulated and computation is dense and compute-bound, low-precision computation can translate into significant system-level gains \citep{quantization5,quantization6,fp8}. However, most existing results are established in the context of large language models, whose standardized architectures and execution patterns differ substantially from those of recommendation models.

In the recommendation domain, prior work has explored low-precision techniques primarily for embedding compression or memory efficiency \citep{low-precision-embedding}. While these studies demonstrate the feasibility of applying low-precision representations to specific components, they typically do not investigate distributional properties or end-to-end inference behavior under low-precision deployment. As a result, the system-level effectiveness of low-precision quantization in modern recommendation models remains underexplored.

\subsection{Recommendation Models and Generative Recommendation}
Recommendation models have traditionally been developed under supervised learning paradigms, with objectives such as click-through rate or conversion rate. Classical industrial architectures commonly adopt multi-stage pipelines, where retrieval, filtering, and ranking are implemented as separate components. These systems typically rely on large embedding tables to represent sparse categorical features, followed by multi-layer perceptrons or feature interaction modules for prediction \citep{reco-system,scale-reco,review-reco,torchrec}. While effective in practice, the multi-stage design inherently constrains downstream models by the capacity and recall limits of upstream stages, and the overall inference pipeline often exhibits heterogeneous and loosely coupled computation patterns.

Recently, generative recommendation has emerged as an alternative paradigm that formulates recommendation as conditional sequence generation \citep{gr}. By integrating retrieval and ranking into a unified generative process, this approach removes strict stage boundaries and enables end-to-end optimization within a single modeling framework. OneRec represents a systematic exploration of this paradigm \citep{onerec}, and subsequent extensions such as OneRec-V2 further scale the architecture through training and structural refinements \citep{onerec-v2,onerec-think,openonerec}. Compared to traditional multi-stage pipelines, generative recommendation models rely more heavily on dense computation paths and unified execution patterns, which can lead to higher hardware utilization during training and inference.

In parallel, a growing body of work has examined the interaction between large language models and recommendation tasks, including empirical studies of LLM-based recommendation capabilities and surveys of emerging design trends \citep{reco1,reco2}. These developments further highlight the convergence between generative modeling and modern recommendation workloads.
\section{Preliminary}

\subsection{Background on Quantization}
Quantization is a technique that represents model parameters and intermediate activations using lower numerical precision in order to reduce computation cost, memory footprint, and data movement during inference. Given a floating-point tensor $\mathbf{x}$, quantization typically involves rescaling its values to match the representable range of a target format, followed by rounding. A generic quantized representation $\hat{\mathbf{x}}$ can be expressed as
\[
\hat{\mathbf{x}} = Q(\mathbf{x}; s) = \mathrm{round}\left(\frac{\mathbf{x}}{s}\right),
\]
where $s$ denotes a scaling factor that controls how the dynamic range of $\mathbf{x}$ is mapped to the target precision. During computation, quantized values are often accumulated in higher precision or combined with appropriate scaling to control numerical error.

Low-precision quantization can be applied to both model weights and activations. Key design choices include the granularity of scaling (e.g., per-tensor, per-channel, or block-wise) and whether scaling factors are determined statically or dynamically at runtime. In LLMs, low-precision formats are typically used in conjunction with higher-precision accumulation for dense linear operations such as matrix multiplication, where most of the computational cost resides.

The effectiveness of quantization depends critically on the numerical properties of the tensors being quantized. In particular, the magnitude and variance of weights and activations determine whether low-precision representations introduce tolerable perturbations or lead to harmful numerical distortion. As a result, understanding the distributional characteristics of a model is an essential prerequisite for assessing the feasibility of low-precision inference.

\subsection{Distribution Characteristics of Model Weights and Activations}

\begin{figure*}[t]
    \centering
    \begin{subfigure}[t]{0.48\linewidth}
        \centering
        \includegraphics[width=\linewidth]{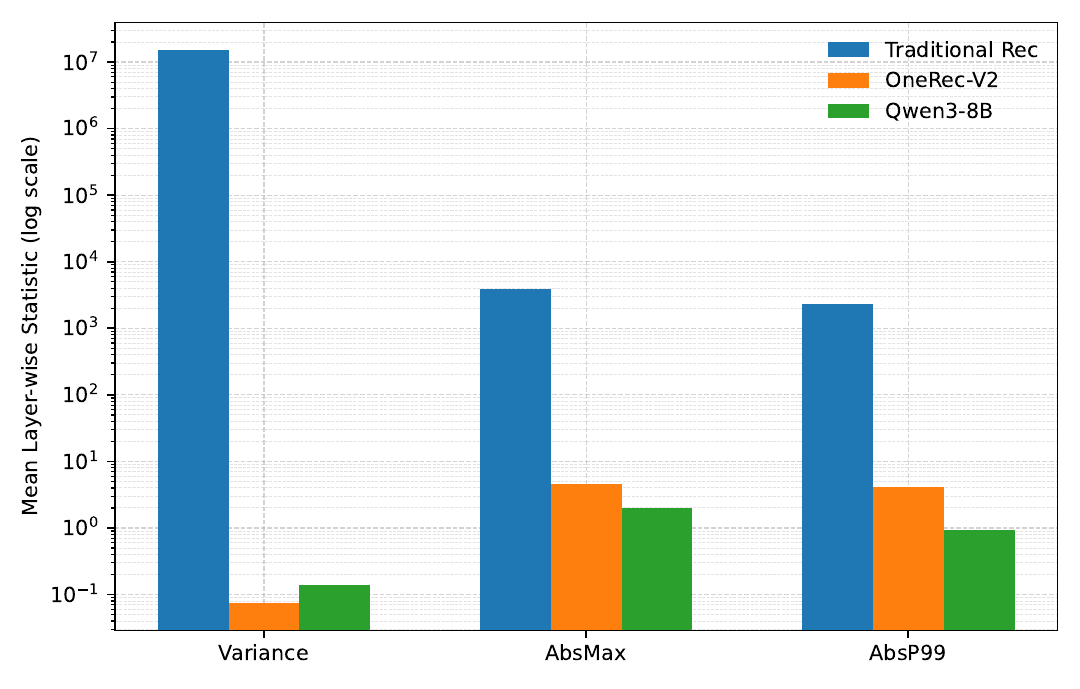}
        \caption{Weight distribution statistics (log scale).}
        \label{fig:dist_weight}
    \end{subfigure}
    \hfill
    \begin{subfigure}[t]{0.48\linewidth}
        \centering
        \includegraphics[width=\linewidth]{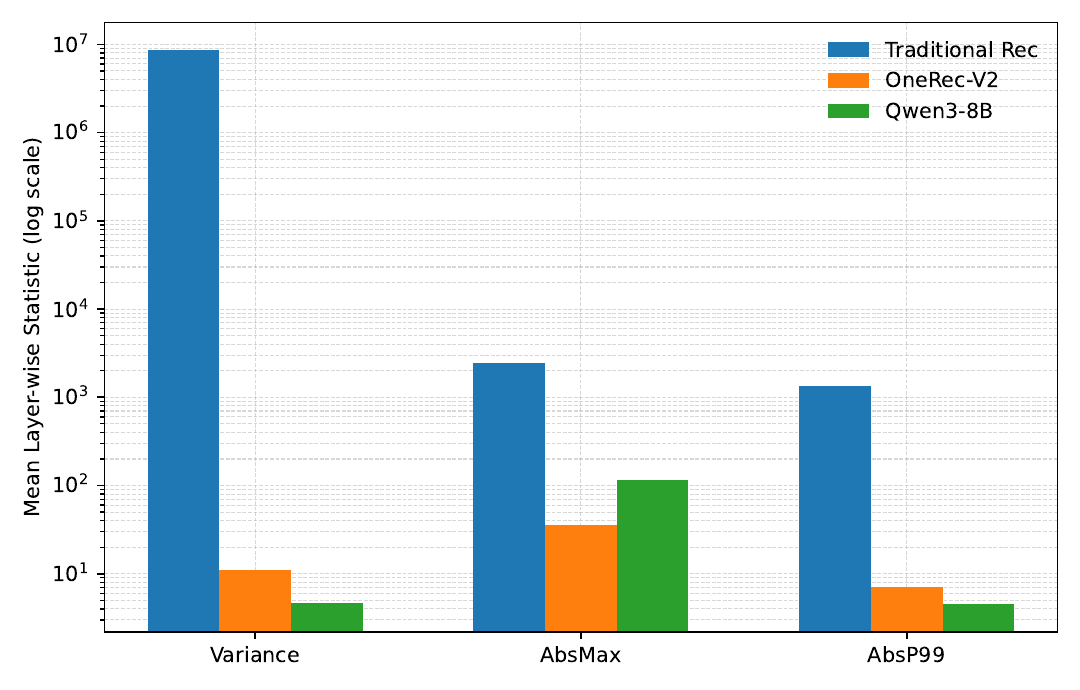}
        \caption{Activation distribution statistics (log scale).}
        \label{fig:dist_activation}
    \end{subfigure}
    \caption{
    Comparison of distribution statistics across different model families, including a classical recommendation model, OneRec-V2, and Qwen3-8B.
    All values are reported as the mean variance, mean absolute maximum (AbsMax), and mean 99th percentile absolute value (AbsP99) across all tensors, and plotted in log scale.
    The classical recommendation model exhibits significantly larger dynamic ranges and variances compared to OneRec-V2 and Qwen3-8B, while OneRec-V2 demonstrates distributional characteristics closer to those of Qwen3-8B.
    }
    \label{fig:distribution_comparison}
\end{figure*}

To understand the feasibility of low-precision inference across different model families, we analyze the empirical distributions of model weights and activations. We compare three representative models: (1) a recommendation model used for fine-grained ranking (snapshot dated 2026-02-04), (2) OneRec-V2 (snapshot dated 2026-02-27), and (3) Qwen3-8B as a representative LLM. For each model, we collect statistics including variance, absolute maximum (AbsMax), and 99th-percentile absolute value (AbsP99), and report their mean values across all tensors. These metrics capture both overall distribution and extreme-value behavior, which are critical factors for quantization.
Figure~\ref{fig:distribution_comparison} summarizes these statistics in log scale.

\noindent \textbf{Traditional recommendation model.}
The traditional recommendation model exhibits extremely large numerical scales in both weights and activations. The mean variance of weights is on the order of $10^{7}$, and the mean AbsMax exceeds $10^{3}$. Activation statistics show a similar pattern, with mean variance also on the order of $10^{6}$ and large extreme-value magnitudes. These results indicate highly dispersed distributions and wide dynamic ranges of the model.

\noindent \textbf{Large language model.}
In contrast, Qwen3-8B exhibits substantially smaller weight statistics. 
The mean variance of weights is about $0.1$, and extreme values is about $2.0$. 
While activation maximum and variance may occasionally reach higher values, their overall scale remains markedly lower than those observed in the traditional recommendation model.

\noindent \textbf{OneRec-V2.}
We can conclude from the figure that OneRec-V2 demonstrates distributional statistics that are much closer to those of the large language model than to the traditional recommendation model. The mean weight variance is below $0.1$. Activation statistics are also orders of magnitude smaller than those observed in the classical recommendation model, with substantially reduced dynamic ranges and dispersion.

These empirical observations indicate that the numerical behavior of weights and activations differs fundamentally across model families. Models with extremely wide dynamic ranges and highly dispersed distributions present greater challenges for coarse-grained scaling, whereas models with more controlled statistics are inherently more compatible with low-precision representations.
\section{Method}

Based on the distributional analysis presented in the previous section, we design a practical low-precision inference solution for OneRec-V2. Our goal is not only to introduce low-precision computation into the model, but also to ensure that such numerical changes translate into meaningful end-to-end system improvements under real serving workloads.
To this end, we adopt a joint optimization strategy that consists of two complementary components. First, we develop a post-training quantization scheme tailored to the numerical and computational characteristics of OneRec-V2. Second, we integrate the quantized model into an optimized inference infrastructure that improves hardware utilization and execution efficiency. The combination of these two components enables low-precision computation to deliver stable and measurable system-level gains.

\subsection{Quantization Strategy}
\begin{figure*}[t]
    \centering
    \includegraphics[width=0.8\linewidth]{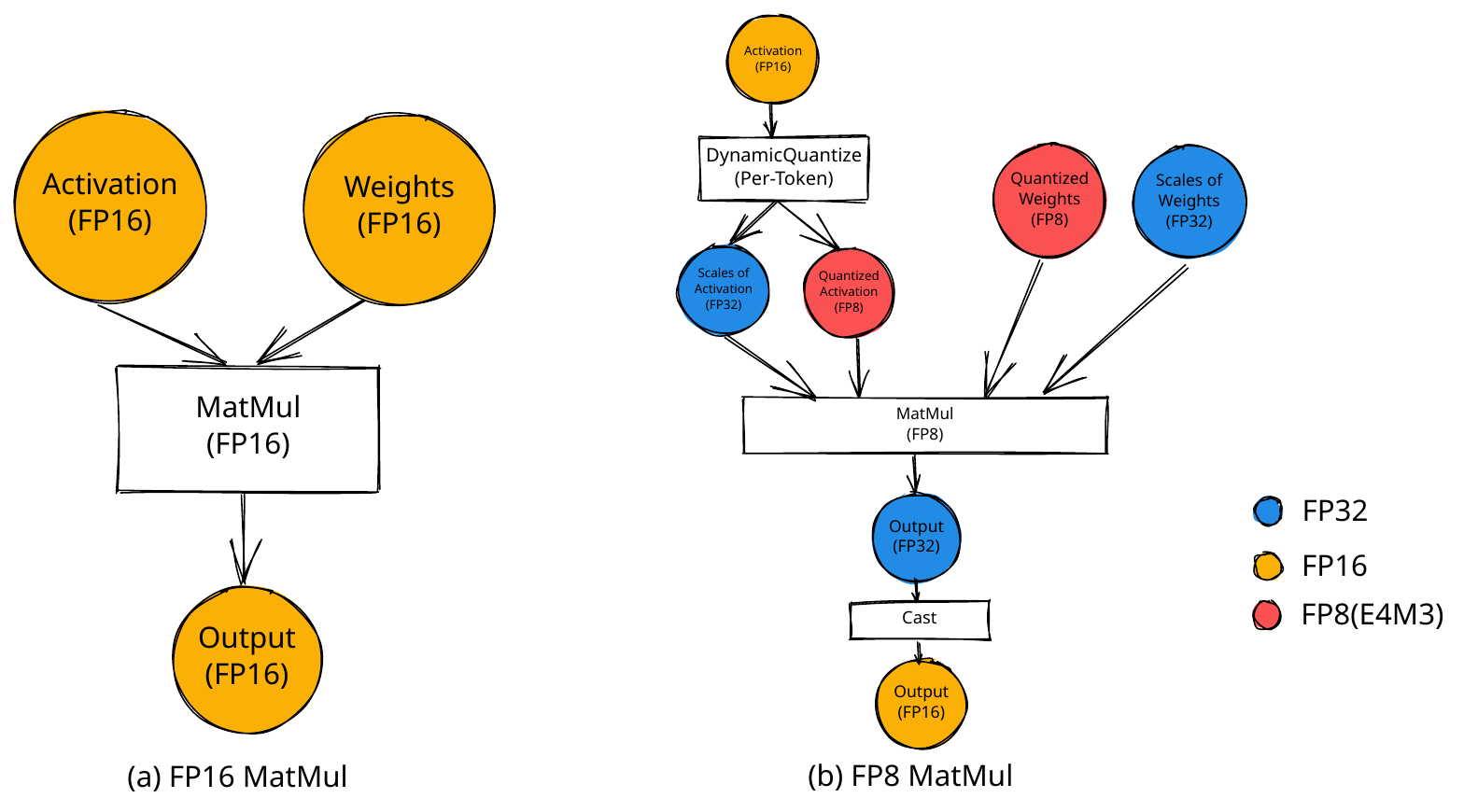}
    \caption{
    Comparison between FP16 and FP8 linear computation. In the FP8 path, inputs are first rescaled and quantized to low precision, followed by low-precision matrix multiplication with FP32 accumulation, and finally cast back to higher precision for subsequent layers. The FP16 path performs matrix multiplication directly in higher precision.
    }
    \label{fig:fp8_linear}
\end{figure*}

We adopt a post-training quantization (PTQ) approach to introduce low-precision computation into the inference stage of OneRec-V2 without modifying the model architecture or training procedure. Quantization is applied only to the most computation-intensive operators, namely the Linear layers (including the qkvo projection layers in Attention and the linear transformations in Dense FFN) and the grouped GEMM operations in Sparse MoE. Other numerically sensitive or less compute-dominant components remain in their original precision to control potential numerical risks.

Figure \ref{fig:fp8_linear} shows the comparison between FP16 matmul and FP8 matmul of Linear layers. In Linear layers, weights are quantized by channel, with scales computed offline from the FP16 parameters, while activations are quantized dynamically by token using runtime-computed scales. 
Matrix multiplication is performed using FP8 TensorCore multiplication with FP32 accumulation, the results are cast back to FP16 before entering subsequent layers. 
For grouped GEMM in MoE, we adopt block-wise quantization, using a $1\times128$ granularity for activations along the last dimension and a $128\times128$ granularity for weights.
This block-aligned design reduces the dynamic range covered by a single scale while preserving the original routing and parallel execution structure, allowing low-precision computation to be integrated without accuracy degradation.
All model weights are pre-quantized and stored in a (FP8 weight, FP32 scale) pair in GPU memory.

Overall, the quantization strategy prioritizes coverage of compute-intensive operators and controls numerical error through structure-aligned scaling and higher-precision accumulation, providing a stable numerical foundation for subsequent system-level inference optimization.

\subsection{Infrastructure Optimization}
In addition to quantization, we optimize the inference infrastructure (RecoGEM) to ensure that reduced-precision computation translates into measurable end-to-end performance gains. 
Instead of relying on a multi-stage conversion pipeline (e.g., PyTorch $\rightarrow$ ONNX $\rightarrow$ TensorRT), we construct the TensorRT execution graph directly using a unified operator library, enabling tighter control over kernel selection, operator fusion, and memory layout.

On top of this execution framework, we implement several operator-level optimizations:
\begin{itemize}
    \item \textbf{Quantization operators.} 
    We implement an efficient per-row quantization operator for activation scaling and a fused low-precision matrix multiplication kernel that integrates quantization and GEMM execution, reducing intermediate memory traffic and improving compute efficiency.

    \item \textbf{TopK optimization.} 
    We replace the default TopK implementation with a radix-based kernel (RadixTopK) and apply kernel fusion and zero-copy techniques to reduce memory movement and improve throughput under large-batch workloads.

    \item \textbf{Attention optimization.} 
    We redesign the attention kernel for the large-batch, short-context setting typical of OneRec-V2, introducing batch-level parallelism and software pipelining to improve scheduling efficiency and hardware utilization.

    \item \textbf{MoE optimization.} 
    For Sparse MoE, we optimize the grouped GEMM execution path using Hopper TMA-enabled kernels and operator fusion, improving TensorCore utilization and increasing the effective model FLOPs utilization (MFU).
\end{itemize}

Together, these infrastructure-level optimizations increase arithmetic intensity and hardware utilization during inference. 
When combined with the quantization strategy described earlier, the optimized inference stack enables substantial latency reduction and throughput improvement under production serving configurations.
\section{Experiments}
In this section, we evaluate the effectiveness of the proposed low-precision inference approach for OneRec-V2. 
We first report offline serving performance, including end-to-end latency and throughput, and analyze the contribution of infrastructure upgrade, quantization, and operator-level optimizations. 
We then present online A/B testing results to verify that the proposed deployment achieves substantial efficiency gains without degrading recommendation quality.

\subsection{Setup}
\noindent \textbf{Model.}
We evaluate a production-scale OneRec-V2 model with a fat-MoE architecture, consisting of approximately 4B backbone parameters with 0.5B activated per token. The model follows a generative recommendation paradigm and is deployed in a single-column short-video recommendation scenario.

\noindent \textbf{Configuration.}
All experiments are conducted with batch size 32 under an online serving configuration. The baseline system performs inference in FP16. The proposed system applies post-training quantization to computation dominant Linear layers in Attention, Dense FFN, and MoE expert modules, while other components remain in FP16.

\subsection{System Performance Evaluation}

We first evaluate the system-level efficiency of the proposed deployment in terms of end-to-end latency and throughput.

The baseline FP16 system achieves an average inference latency of 139\,ms and a throughput of 205. 
After migrating to the optimized inference infra and enabling low-precision computation, latency is reduced to 70\,ms, corresponding to a 49\% reduction. 
Throughput increases from 205 to 394, yielding a 92\% improvement over the baseline.

\begin{figure}[t]
    \centering
    \includegraphics[width=1 \linewidth]{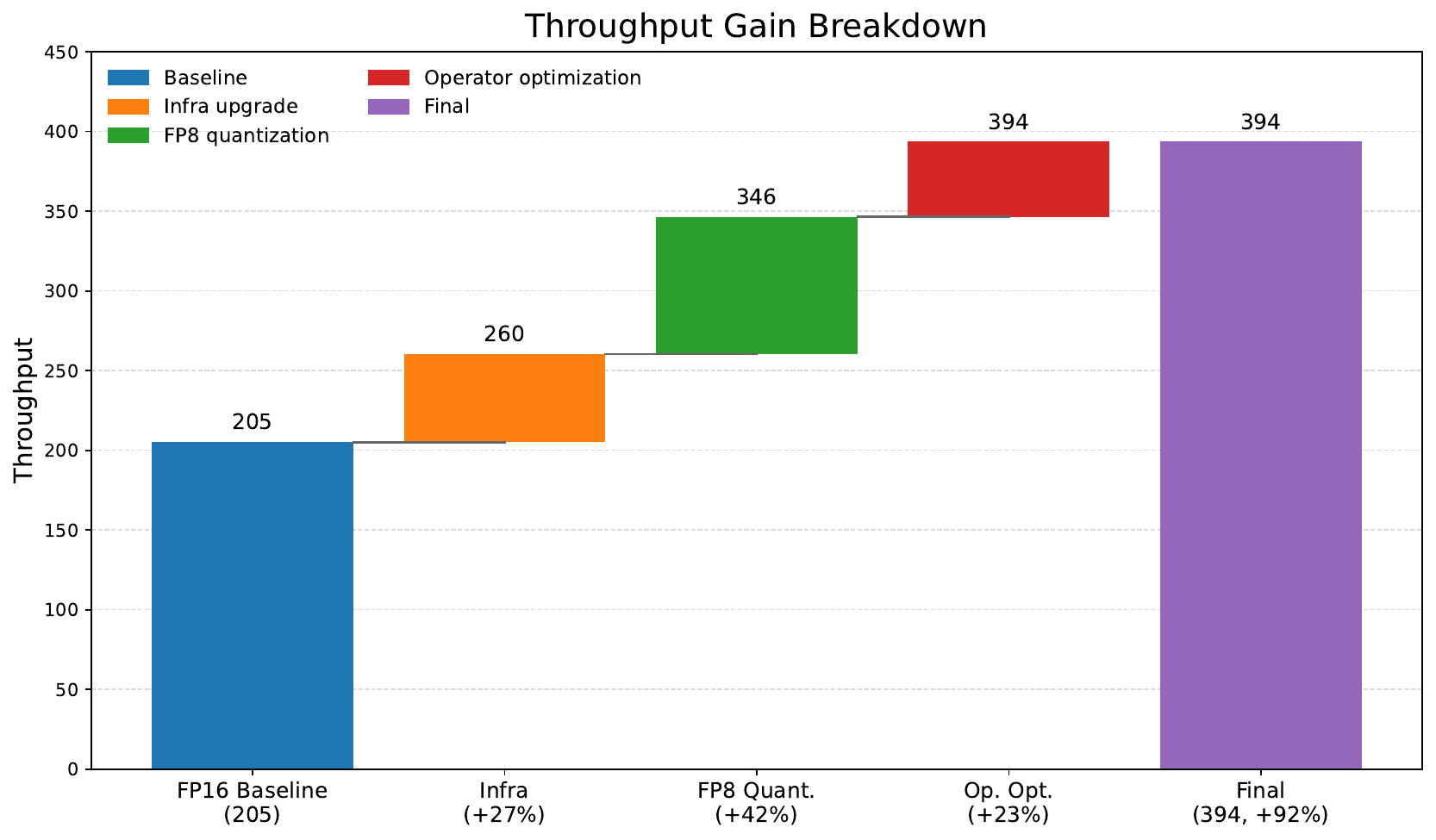}
    \caption{
    Throughput gain breakdown.
    Starting from a baseline throughput of 205, the migration to the optimized inference architecture contributes a 27\% improvement, enabling FP8 quantization on Linear and MoE operators contributes an additional 42\%, and operator-level optimizations contribute a further 23\%. 
    These additive improvements result in a final throughput of 394, corresponding to a 92\% end-to-end increase over the baseline.
    }
    \label{fig:throughput_breakdown}
\end{figure}

To understand the sources of these gains, we further decompose the throughput improvement into three components, as shown in Figure~\ref{fig:throughput_breakdown}. 
Infrastructure upgrade contributes a 27\% increase relative to the baseline, enabling low-precision computation on Linear and MoE operators contributes an additional 42\%, and operator-level optimizations contribute a further 23\%. 
These additive improvements collectively account for the observed 92\% end-to-end throughput gain.

Overall, the results indicate that infrastructure upgrade, low-precision quantization, and operator-level optimizations each provide measurable performance gains. 
While quantization contributes a substantial portion of the throughput improvement, the full 92\% increase is realized through their combined and complementary effects.

\subsection{Online A/B Evaluation}

\begin{table}[t]
\centering
\caption{
Online A/B results under FP8 inference compared to the FP16 baseline.
}
\label{tab:ab_results}
\begin{tabular}{lcc}
\toprule
\textbf{Metric} & \textbf{Kuaishou} & \textbf{Kuaishou Lite} \\
\midrule
App Stay Time & +0.047\% & -0.018\% \\
Watch Time & +0.044\% & -0.157\% \\
Video View & +0.073\% & -0.041\% \\
Like & +0.085\% & -0.102\% \\
Follow & +0.020\% & -0.450\% \\
Comment & +0.805\% & +0.261\% \\
Collect & +0.430\% & -0.353\% \\
Forward & +0.313\% & +1.047\% \\
\bottomrule
\end{tabular}
\end{table}

To evaluate the impact of low-precision inference on recommendation quality, we conduct online A/B testing in production environments over a one-week period.
Table~\ref{tab:ab_results} shows that core metrics remain stable under the optimized inference configuration in both Kuaishou and Kuaishou Lite. 
No consistent performance degradation is observed across key indicators.
Overall, the results confirm that low-precision inference can be deployed without compromising recommendation effectiveness in real-world settings.

\subsection{Analysis and Discussion}
The experimental results support two main observations.
First, low-precision inference can deliver substantial system-level acceleration in generative recommendation models whose computation is dominated by dense linear operations and MoE experts. The nearly 2$\times$ throughput improvement demonstrates the practical value of reduced-precision computation in large-scale recommendation serving.
Second, despite applying lower precision to key computation paths, online metrics remain stable. Combined with the distribution analysis in Preliminary, this indicates that the weight and activation statistics of OneRec-V2 are sufficiently controlled to tolerate quantization noise.

Overall, the results demonstrate that low-precision inference is both feasible and beneficial for large-scale generative recommendation models under real-world production constraints.
\section{Conclusion}

In this work, we study low-precision inference in the context of generative recommendation and demonstrate its practicality on a production-scale OneRec-V2 model. 
We show that the feasibility of quantization in recommendation models depends not only on numerical formats, but also on model architecture, training paradigm, and inference execution characteristics. 
Compared to traditional recommendation models, OneRec-V2 exhibits more controlled weight and activation statistics and a more compute-intensive execution pattern, making it more amenable to low-precision computation.

Building upon these properties, we integrate post-training quantization with an optimized inference infrastructure and operator-level improvements. 
Our experiments show that each component contributes positively to system efficiency, and their combination leads to substantial end-to-end latency reduction and throughput improvement, while preserving online recommendation quality under real-world applications.

These results suggest that as recommendation models evolve toward dense, generative architectures, low-precision computation can become a practical and scalable tool for improving serving efficiency. 
More broadly, optimization techniques developed for large language models can be effectively adapted to modern recommendation models when architectural and system-level considerations are jointly addressed.

\section*{Limitations}

\noindent \textbf{No exploration of lower precisions.} This work focuses on FP8 inference as a practical trade-off between efficiency and deployment stability, and does not investigate more aggressive low-precision settings such as INT8, FP6, FP4, or mixed lower-bit schemes. As a result, the current study does not reveal the full accuracy-efficiency frontier of generative recommendation models, and it remains unclear whether further compression can be achieved without noticeable quality degradation.

\noindent \textbf{High infrastructure and requirements.} The proposed solution relies on substantial infrastructure support and system-level customization, including hardware capability, TensorRT-based deployment, operator-level adaptation, and tailored optimization for key modules such as Linear and MoE. These requirements may limit the reproducibility and portability of the approach, especially in production environments that lack advanced inference stacks or sufficient engineering resources.

\noindent \textbf{Limited coverage of generative recommendation models.} Our conclusions are drawn from experiments on OneRec-V2, and we do not study a broader set of generative recommendation architectures. Therefore, it is still unclear to what extent the observed quantization properties and deployment benefits generalize across different model designs, training paradigms, and recommendation scenarios. Broader validation on more generative recommendation models is needed in future work.

\section*{Acknowledgments}
We thank Haiping Xu, Xu He, Chenhui Wang, and Jin Ouyang for their contributions in training infrastructure.
\bibliography{custom}

\appendix

\section{Example Appendix}
\label{sec:appendix}

This is an appendix.

\end{document}